\documentclass{aa}

\begin{document}

\title{Iron line afterglows: how to produce them}

\author{D. Lazzati\inst{1,2}, S. Campana\inst{1} \and G. Ghisellini\inst{1}}
\authorrunning{Lazzati et al.}

\institute{Osservatorio Astronomico di Brera, Via Bianchi 46, I--23807
Merate (Lc), Italy
\and
Dipartimento di Fisica, Universit\`a degli Studi di Milano,
Via Celoria 16, I--20133 Milano, Italy}

\mail{lazzati@merate.mi.astro.it}

\maketitle

\begin{abstract}

We discuss how a powerful iron line emission can be produced 
if $\sim 1-5$ iron rich solar masses are concentrated in the close 
vicinity of the burst. Recombination, thermal and fluorescent 
reflection are discussed. We find that recombination suffers the
high Compton temperature of the plasma while the other two scenarios
are not mutually exclusive and could account for the claimed 
iron line detected in two afterglows.

\end{abstract}

\section{Introduction}
Piro et al. 1998 and Yoshida et al. 1998 (this volume) 
report the detection of an 
iron emission line in the X--ray afterglow spectrum of two $\gamma$--ray
bursts, GRB~970508 and GRB~970828.
The detection of strong iron emission lines would unambiguously point towards 
the presence of a few per cent of
iron solar masses concentrated in a compact region in the vicinity of 
the burster.
The simplest way to account for this is to assume the presence of
a very young supernova remnant.
Thus the presence of such a line in the X--ray afterglow spectrum
would represent the ``Rosetta Stone'' for unveiling the burst progenitor.
There are three main classes of  models for the origin of 
GRB: Neutron Star--Neutron Star (NS--NS) mergers
(Eichler et al. 1989), Hypernovae (Paczy\'nski 1998) and Supranovae 
(Vietri \& Stella 1998). 

X--ray line emission following GRB events has been recently 
discussed in the Hypernova scenario by Ghi\-sellini et al. 1998 and \cite{boe98}. 
None of these works predict, with reasonable assumptions on the burst 
surroundings, iron lines strong enough to be detectable during the
X--ray afterglow. Moreover, the line emission should last over a time--scale
of years given the size of the emitting nebula.
On the contrary, the Supranova accounts in a quite natural way for the
presence of a massive remnant in the close vicinity of
the bursting source.

Here we discuss three possibilities for the iron line origin:
$i)$ recombination of the supernova shell iron atoms, photoionized by
the burst photons;
$ii)$ thermal emission of the shell, which, once heated by the burst emission, 
produces an iron blended line (mainly due to FeXXV and FeXXVI);
$iii)$ fluorescence of a very dense, compact and relatively cold supernova shell
observed in reflection.

We assume that $\sim 1-5$ iron rich solar masses
are concentrated in the burst surroundings. 
Some general constraints about the required mass and the size of the 
emitting region are discussed in Ghisellini et al. 1998 (this volume), whereas
a more detailed description can be found in Lazzati et al. (1998).
The cosmological parameters will be set throughout this letter to
$H_0 = 65$~km~s$^{-1}$~Mpc$^{-1}$, $q_0=0.5$ and $\Lambda=0$.

\section{Emission mechanisms}

\subsection{Recombination}

If the plasma can remain cold and dense enough and if the recombination
and ionization times are comparable, a strong iron line can be obtained 
through recombination during the burst with a reasonable amount of iron.
A recombination time of almost $10^{-2}$~s is needed to produce a line 
visible during the afterglow.

The recombination time of an hydrogenic ion of atomic number $Z$
in a thermal plasma is $t_{rec} = (\alpha_r \, n)^{-1}$ (Verner \& Ferland 1986),
where $n$ is the electron density and $\alpha_r$ is the temperature
dependent recombination coefficient (e.g. Seaton 1959).
For a relatively cold and dense iron plasma with $T=10^4$~K,
$t_{rec} \sim 0.14 \, n_{10}^{-1}$~s while, for
$T=10^8$~K, $t_{rec} \sim 10^{2} \, n_{10}^{-1}$~s.
Hence, while a cold plasma could have the necessary
recombination rate to ensure the needed fast photon production,
this mechanism would be almost completely damped at higher
temperatures with reasonable values for the electron density.

The main problem with this interpretation is that the temperature
of the plasma absorbing a sizeable fraction of the burst energy
is bound to be large, unless an extreme large density enhances
the radiative cooling rates.
In fact the Compton equilibrium temperature (assuming typical
burst high energy spectra) is of the order of a few times $10^8$ K.
Note that the scattering optical depth $\tau_T$ is bound to be
in the range 0.1--1 since a sizable fraction of the burst energy must 
be absorbed without an excessive smearing of the emission lines.

We conclude that the case of a strong iron line due to repeated 
photoionization and recombination events during the burst emission 
faces the problem of a temperature too large to ensure the required 
fast recombination rate.

\subsection{Thermal emission from the surrounding shell}

We assume for simplicity that the shell is homogeneous
and compact and that it is heated up to $T=10^8\,T_8$~K by the burst photons.
We must require that $\tau_T \simeq 0.1 \div 1$,
implying a shell radius $R\ge 8\times 10^{15}(M/M_\odot)^{1/2}$.
In this conditions the shell would emit a broad band brems\-strah\-lung 
continuum with several emission lines overlaid (\cite{ray77}), the most 
relevant being the 6.7 keV iron blend. Analogous spectra are observed
in cluster of galaxies (\cite{sar88}), 
but the higher iron abundance expected in a 
supernova shell would enhance line emission.

The equivalent width (EW) of the line in a solar abundance plasma
has been carefully computed by Bahcall \& Sarazin (1978; see in particular 
their Figure 1) and ranges from several tens of eV at high ($5\times 10^8$~K)
temperatures to $\sim 2$~keV at $2.5 \times 10^7$~K. 
A very weak line is expected for temperature lower than $5\times10^6$~K.
For temperatures larger than $5\times10^7$~K the EW dependence on 
temperature can be reasonably approximated as a power law.
Assuming an iron abundance 10 times solar we have:
\begin{equation}
\hbox{EW}(T) \simeq 3.8 \, T_8^{-1.9}\quad {\rm keV}\quad (T_8 \geq 0.5)
\label{ewt}
\end{equation}
Considering the brems\-strah\-lung intensity at 6.7~keV, 
we obtain a line luminosity of:
\begin{equation}
L_{Fe} \simeq 8 \times 10^{42}  \exp\left(-{0.8\over T_8}\right)
\, \left({M \over M_\odot}\right)^2  
V_{48}^{-1}  T_8^{-2.4} \; \hbox {erg s$^{-1}$}
\label{llin}
\end{equation}
Therefore a shell of $M \sim 5 M_\odot$, typical
for many type II SN (see \cite{ray84,wei88,woo88,mcc93}),
at a temperature slightly below $10^8$ K can produce a line flux of 
$10^{-13}$ erg cm$^{-2}$ s$^{-1}$ for $z=1$ bursts.
The EW with respect to the underlying brems\-strah\-lung radiation would be a few
keV, but any other emission component (e.g. afterglow emission) 
would decrease the line EW.
The line emission process can be stopped after about one day,
if the afterglow photons enhance the
plasma cooling via inverse Compton, lowering the temperature down to
less than $10^7$~K. Line emission can also be quenched by the reheating 
produced by the incoming fireball.

\subsection{Reflection}

In Seyfert galaxies we see a fluorescence 6.4 keV iron line produced
by a relatively cold ($T<10^6$ K) accretion disk, illuminated
by a hot corona, which provides the ionizing photons 
(e.g. \cite{ros93}).
In this case we need a scattering optical depth 
$\tau_T>1$ of the fluorescent material and 
a size large enough to allow the line being emitted even $\sim$~one day
after the GRB event (i.e. $R > 10^{15}$ cm).

In this model the emission line is produced only during the burst event,
but in the observer frame it lasts for $R/c$.
The observed luminosity of the Compton reflection
component is equal to the $\sim$~10\% 
of the absorbed energy, divided by the time $R/c$: 
$L\sim 3\times 10^{45} \,E_{abs, 51}/R_{15}$
erg s$^{-1}$.
The luminosity in the iron line (see e.g.~\cite{mat91}) is $\sim 1\%$ 
of this, times the iron abundance in solar units.
Therefore the reflection component 
can contribute to the hard X--ray afterglow emission, 
and the iron line can have a
luminosity up to $3\times 10^{44}$ erg s$^{-1}$, corresponding to fluxes
$\sim 10^{-13}$ erg cm$^{-2}$ s$^{-1}$ for a $z=1$ burst.

\section{Conclusions}

We discussed three possible mechanisms for the production of
a strong iron line, visible during the X--ray afterglows of GRB.
All mechanisms require the presence of a large amount of iron in a compact
region.
The remnant of a supernova exploded a few months before the burst
matches the required matter and size.

We find that the multiple ionization and recombination scenario
has difficulties in reconciling the low temperature required to have
a fast recombination with the large heating due to the burst flux.
On the other hand, we believe that the other two alternatives (i.e.
thermal emission and reflection) are promising and not
mutually exclusive.

In the reflection scenario, afterglow spectra should show the typical 
hardening of the spectrum above a few keV, and the line duration
should be short.

\end{document}